\documentclass[10pt,a4paper]{article}
\usepackage{amsmath}
\usepackage{latexsym}
\usepackage{theorem}
\usepackage[all]{xy}

\include{amslatex}

\textheight
25 cm \textwidth 17 cm \oddsidemargin -5 mm \topmargin -20 mm
{\theorembodyfont{\rmfamily}
}
{\theorembodyfont{\rmfamily}
}

\numberwithin{equation}{section}

\begin{document}
\begin{title}
{\Large {\bf Emergent phenomenology from deterministic Cartan-Randers systems}}
\end{title}
\maketitle
\author{
\begin{center}

Ricardo Gallego Torrom\'{e}\footnote{email: email: rigato39@gmail.com. Currently at the
 Departamento de Matem\'atica, Universidade Federal de S\~ao Carlos, Brazil.}\\
Riemann Center for Geometry and Physics\\
Institute of Theoretical Physics,
Leibniz Universit\"{a}t Hannover\\
\end{center}}

\begin{abstract}
We consider specific {\it deterministic quantum models} of Cartan-Randers type and show how a dualized abelian gauge symmetry, {\it diffeomorphism invariance}, the Principle of Inertia, reversibility of phenomenological dynamics, {\it maximal acceleration} and {\it maximal propagation speed for matter} arise from such models in an phenomenological description. Two particular predictions are highlighted: the first is the value of maximal acceleration is universal and has the value of order $10^{52} m/s^2$. The second prediction is that gauge symmetries in phenomenological models must come dualized and containing an abelian sector.
\end{abstract}

\section{Introduction}
\
  Besides the spectacular success of quantum mechanics in describing the atomic and sub-atomic phenomena,
  the failure in finding a consistent quantum theory of
  gravity and the existence of the cosmological constant problems are a quite puzzling facts.
  To find a consistent quantum description of gravity is recognized as one of the most relevant in the foundation of physics and although one should find fundamental problems in the search of quantum gravity, the most shared opinion is that fundamental theories must be quantum theories and this must be the case for a more deeper theory of gravity. 

However, in view of the elusive {\it consistent theory of quantum gravity} and the still problematic issues in the interpretation of the foundations of the quantum theory, it is reasonable to pondered if such programmatic direction of research towards a quantum theory of gravity is the right one. Thus, one could arguably doubt if the quantum formalism is the finest way of description of reality and if we have a correct understanding of what is gravity.

  Several alternative theories to quantum mechanics are recognized under the common nomination of {\it emergent
  quantum mechanics}. They share the idea that the
  quantum formalism arises from an underlying more fundamental
  theory. We cannot make full justice here to the whole contributions to this subject, but let us mention the emergent theories developed in \cite{Adler, Blasone, Blasone2, Nelson, Elze, Hooft,  Smolin} and other recent contributions. Without being  {\it hidden variable theories} in the usual sense,  these emergent frameworks aim to reproduce the quantum mechanics formalism from
  a more fundamental deterministic formalism
  and to obtain observable deviations from the standard quantum
  mechanical predictions. A common characteristic of emergent quantum
  models is that the fundamental degrees of freedom are not necessarily
  the same than the degrees of freedom associated with sub-atomic, atomic, molecular or classical systems, but are described
  by different dynamical variables. Thus, different emergent frameworks  differ on the
  choice of fundamental variables and their dynamics. Moreover, there are mechanisms to skip the validity of Bell's inequalities.

The deterministic models that we are interested here are related with the ones introduced by G. t' Hooft \cite{Hooft}. To introduce such models in a short way, let us describe 
t' Hooft's proposal of deterministic quantum mechanics models as based 
on three fundamental assumptions:
 \begin{itemize}
\item H.1 The ontological degrees of freedom are deterministic  dynamical systems, described by a discrete system of first order ordinary differential equations.

\item H.2 These dynamical systems admit a {\it quantum description} as eigenvectors
of an {\it affine Hamiltonian} in  the {\it momentum operators}.

\item H.3 The dynamics of the {\it ontological degrees of freedom} is dissipative: several
states (actually a large number of them) evolve to the same limit cycles in a finite evolution time.

\end{itemize}
For  dissipative systems with a large number of degrees of freedom a
 probabilistic description is useful, due to the difficulties to follow
 the details of the evolution. This is the underlying reason (in the contest of deterministic quantum mechanics) for the
 success of probabilistic methods in the description of quantum mechanical systems that otherwise are deterministic although extremely complex systems.

A fundamental difficulty associated with 't Hooft's deterministic models
is the relation between the degrees of freedom  at the Planck scale with the
degrees of freedom at subatomic, atomic and classical scale. However,
it has been shown by 't Hooft that some quantum field and string
models can be interpreted as deterministic quantum models \cite{HooftBosonicquantumtheory, Hooftstrings}.
This makes the expectation that the same is true for other relevant models.

A second difficulty of deterministic quantum mechanics is that the associated Hamiltonian operators are not bounded from bellow (since the higher order in
momentum operators is one). The mechanism proposed by 't Hooft involves a dissipative dynamics, somehow correlated with gravity. However, gravity could be itself an emergent phenomena, not really present at the Planck scale. In this case, it is difficult to appeal to gravity for the ingredient producing the loss of information.

In the search for a framework where the above difficulties are resolved in a natural way we assume that the following assumptions hold for deterministic quantum mechanics models:
\begin{itemize}

\item A.1. They are deterministic models: for the ontological degrees of freedom,  the initial conditions provides (at least theoretically) knowledge of the full evolution of the system. The models are described in first approximation by ordinary first order differential equations\footnote{A sharper description of the time evolution for the ontological degrees of freedom is through discrete finite difference equations and that macroscopic time is discrete. Therefore, the use of a continuous time parameter $\tau$ will not affect the considerations made in this work.}.

\item A.2. The following locality condition holds: given a system $S$ and its environment $E$, there is a small neighborhood $U$ with $S\subset \,U\subset E\cup S$ such that only $U$ acts directly on $S$.

\item A.3. There is some causality condition: there is a maximal, universal speed for the ontological degrees of freedom. The simplest assumption is that such maximal speed is the speed of the light in vacuum $c$.
    
\item A.4. The systems evolves towards stable limit cycles, where the system stays until the environment $U$ perturbs the system $S$.

\end{itemize}
Although some of the above assumptions are not independent between each other, we will consider them as axioms for our models. This system of axioms is quite similar to Hooft's axioms, being the main difference the existence of a local action as determined by $A.2$  and a notion of causality, as determined by axiom $A.3$. Otherwise, the axioms $A.1, A.4$ are in correspondence with the axioms $H.1, H.3$.

A direct consequence of the assumptions $A.2, A.3$ is the existence of a {\it maximal universal acceleration}. Since $U$ is the maximal neighborhood that could affect directly $S$, there is associated a minimal length. For minimal systems $S$, the length is also universal minimal.
Then the  maximal work as a  result of such action is
\begin{align*}
L_{min} m \,a \sim \delta {m}\,v_{max}^2 ,
\end{align*}
where $a$ is the value of the acceleration in the direction of the total exterior effort is done.
Since  the speed must be bounded, $v_{max}\leq
c$. Also, the maximal work produced by the system on a point particle is
 $\delta {m}\simeq \, -m$. Thus, there is a bound for the value of the acceleration,
\begin{align}
a_{max}\,\simeq \frac{c^2}{L_{min}}.
\label{maximalacceleration}
\end{align}

A framework for dynamical systems with maximal acceleration and maximal speed is provided by
 {\it deterministic Cartan-Randers systems}\footnote{In reference \cite{Ricardo06} such dynamical systems were called {\it deterministic Finslerian models}. We have adopted here the most appropriated notation of {\it deterministic Cartan-Randers systems, an specific type of Cartan space}  \cite{MironHrimiucShimadaSabau}.}
(DCRS) \cite{Ricardo06}. {\it Deterministic Cartan-Randers systems}
are based on a general correspondence between first order dynamical systems and
{\it Cartan-Randers  metrics} (see \cite{MironHrimiucShimadaSabau} for a definition of such geometric spaces)
defined on a convenient cone of the cotangent bundle $T^*TM$. In a DCRS, the Hamiltonian
operator is averaged on a submanifold (not necessarily compact) of the
cotangent space and also is averaged on the  {\it time direction}, that corresponds to a {\it symmetrization} of an otherwise time asymmetric evolution. The averaged is interpreted as a dynamical {\it limit cycle state}, such that the {\it phenomenological states} are zero modes of the total averaged Hamiltonian; this solves the un-boundeness of the Hamiltonian problem. Moreover, as a
consequence of the non-degeneracy condition of the Cartan-Randers metric,
the velocity and acceleration of the degrees of freedom at the Planck's
scale are bounded.

In this paper it is shown how the quantum mechanical Principle of Inertia and invariance under diffeomorphism emerges from DCRS. Following \cite{Ricardo06}, we discuss the symmetries of DCRS  with a discrete (although large) number of real degrees of freedom.
We show how the automorphism group leaving invariant an {\it ontological state},
describing a full state of microscopic degrees of freedom at the Planck scale, determines an abelian $U(1)\times U(1)$ gauge symmetry. Based on this fact, we argue that the {\it effective description at macroscopic scales} of DCRS are abelian gauge models. Moreover, one can associate to the system  Lorentz invariant {\it effective Lagrangians}. In changing from ontological to effective variables, the notions of {\it irreversible time}, ergodicity and information loss dynamics in DCRS play a fundamental role. Finally, a short discussion of the framework  is presented, indicating how our approach is related with 't Hooft's framework as well as possible extensions to Yang-Mills models and its relation with gravity. In view of the methods and theory developed in this work, we suggest that gravity is a macroscopic effect, that should not appear at the Planck scale. Moreover, two particular predictions are highlighted: the first is the value of maximal acceleration is universal and has the value of order $10^{52} m/s^2$. The second prediction is that gauge symmetries in phenomenological models must come dualized and containing an abelian dualized $U(1)\times U(1)$ sub-group.

\section{Deterministic Cartan-Randers systems}
\
A deterministic Cartan-Randers system is a dynamical system defined by the following elements.
First, there is a {\it configuration manifold} $M$ and its tangent bundle $T M$. Each point $u\in TM$ describes a possible configuration event of the collection of the {\it ontological degrees of freedom} of the full physical system. We assume that the ontological degrees of freedom have a characteristic dimension of length very small compared with the usual scales involved in the Standard Model and that its reaction to the action of an {\it external system} is done in a characteristic time much shorter than the usual characteristic time involved in the Standard Model. It is assumed that such scales are the Planck's length and Planck's time. 

The dimension of $TM$ is $\tilde{N}=dim (TM)$. There is also a {\it space-time model $four$-manifold} $M_4$. The natural number $\tilde{N}=8N$ will be very large compared with the natural number $p=2\,dim(M_4)=8$.
The degrees of freedom of the model are smooth functions describing the motion of {\it identical molecules} of a {\it classical  gas} and for each molecule $k$ there is a configuration manifold $M^k_4\cong M_4$ as a configuration space where $\xi$ can live. Then the number $p=8$ corresponds to four space-time coordinates $(\xi^0,\xi^1,\xi^2,\xi^3)$  for the point $\xi\in\,M^k_4$ and four independent velocities coordinates
$T_\xi M^k_4\ni(\dot{\xi}^0,\dot{\xi}^1,\dot{\xi}^2,\dot{\xi}^3)$. The configuration space for a classical gas of point particles is the manifold $M$ of the form
\begin{align}
TM\cong \sum^N_{k=1}\,\oplus\, TM_4
\end{align}
and this is the prototypical dynamical system that we consider.

In the simplest version, the model manifold $M_4$ is endowed with a Lorentzian  metric $\eta_4$. In this paper, the metric $\eta_4$ will be a back-ground, non-dynamical structure. Then there is a pseudo-Riemannian metric $\eta^*_S$ defined on $TM_4$ (the Sasaki-type metric), which is a lift\footnote{Note that in order to define the lift  of $\eta^*_S$ to $TM_4$ what we essentially need is a connection on $TM$. For a Hamiltonian system, let us consider $\tilde{H}\in\,\mathcal{F}(T^*M)$ be the Hamiltonian function and $\omega$ the symplectic form. Then under some assumptions the $1$-form $(\omega(dH,\cdot))^*\in \,\Gamma \, T^*TM$ determines a spray, that allows to define a connection on $TM$.} of the metric $\eta_4$ on $M_4$ to $TM_4$. The dual metric of $\eta^*_S$ is the pseudo-Riemannian metric $\eta_S$.
The Sasaki-type metric $\eta_S$ allows to define the pseudo-Riemannian metric
\begin{align}
\eta=\sum^N_{k=1}\oplus \eta_S
 \end{align}
 on the $8N$-dimensional manifold $TM$.

The Hamiltonian function for a DCRS is related with the geometric notion of Cartan-Randers space. Let $TM$ be as before and consider a time-like vector field $\beta \in \Gamma TTM$ such that
\begin{align}
|\eta(\beta,\beta)|<1.
\label{boundenesscondition}
\end{align}
 Then the Cartan-Randers function $F^*$ is defined to be
\begin{align}
F^*:\mathcal{C}\to {R^+},\quad
(u,p)\mapsto F^* (u,p)=\alpha(u,p)+\beta(u,p).
\label{corandersmetric}
\end{align}
$F^*$ is defined on the open convex cone  $\mathcal{C}\hookrightarrow T^*TM$  such that $\alpha(u,p)=\sqrt{-\eta^{ij}(u)\,p_i \,p_j}$ is real,
where $u \in TM$ and $ p\in { T^*}_u TM$; the condition \eqref{boundenesscondition} implies that $F^*$ is positive definite in the convex cone $\mathcal{C}$.
For a fixed point $u=(x,y)\in TM$, the beta term is given by the expression
\begin{align}
\beta (u,p)=\beta ^i(u)p_i.
\label{betacomponent}
\end{align}
where $\beta^i(u)$ are the components of the vector $\beta$, that acts on the $1$-form $p\in \mathcal{C}$.
 The fundamental tensor $g$ is given by the vertical Hessian of the function
$(F^*)^2$ in natural coordinates $\{(u^i,p_i),\,i=1,...,4N\}$,
\begin{align}
g^{ij}(u,p)=\,\frac{1}{2}\,\frac{\partial^2 (F^*(u,p))^2}{\partial p_i\partial p_j}, \quad i,j=1,...,8N.
\label{fundamentaltensor}
\end{align}
 A Cartan space is a pair $(\tilde{M},F)$ where $\tilde{M}$ is a smooth manifold with $F:T^*\tilde{M}\to R^+$ smooth on $T^*\tilde{M}\setminus \{0\}$ such that it is homogeneous of degree one on the momentum variables and the vertical Hessian \eqref{fundamentaltensor} is positive definite.
Therefore, the function \eqref{corandersmetric} defines a Cartan space of Randers type on $TM$, instead than on $M$. 
The requirement that $F^*$ is a Cartan-Randers function is justified by a direct relation between Cartan structures and Hamiltonian systems (see for instance \cite{Syngespecial1972} and \cite{MironHrimiucShimadaSabau}, {\it Chapter} 5 and 6).  The fundamental fact in such relation is that the function $F^*$ has associated a geodesic flow and a Hamiltonian flow and that both flows coincide on their projections to $M$. To prove such identification the essential requirement is the non-degeneracy of the fundamental tensor \eqref{fundamentaltensor}. 

The space of Cartan-Randers structures $\mathcal{F}^*(TM)$ on $TM$ plays a fundamental role in the formulation of DCRS  since it has associated geometric flows. Let us introduce a compact time parameter $t\in\,[0,1]$ that is the parameter in a geometric evolution of the Cartan-Randers structures in $\mathcal{F}^*(TM)$. The existence of such geometric flows follows in analogy to the geometric flows in Riemannian geometry (in particular, to the mean curvature flow). However, we will not need here  the details of such geometric formulation of the geometric flow in $\mathcal{F}^*(TM)$, since we only need for our purposes an averaged operation of geometric structures on $TM$ and the corresponding homotopy transformation \cite{Ricardo}. 

Given the Cartan-Randers space
 $(TM, F^*)$, it is possible to define an averaged
Riemannian metric $h$ \cite{Ricardo} by averaging the
fundamental tensor $g$ in each co-tangent space $ {T_u M }$. Thus the averaging of $F^*$ is defined by the averaging of the metric coefficients $g_{ij}$ on the cone $\mathcal{C}_u$ of $u\in TM$,
\begin{align*}
h_{ij}(u)=\langle g_{ij}(u,p)\rangle \,:=\,\frac{1}{\int_{\mathcal{C}_u}\,dvol_u(p)}\int_{\mathcal{C}_u}\,dvol_u(p)\,g_{ij}(u,p),\quad i,j=1,...,8N.
\end{align*}
Although the cone $\mathcal{C}_u$ is not compact, the volume form $vol_(p))$ is chosen such that the integrals are defined. We also make the hypothesis that the volume function $\,dvol_u(p)$ is invariant under the required symmetries of the system.

The $t$-evolution towards the average state is a dissipative evolution of the Cartan-Randers spaces,
\begin{align}
U_t: (T^*TM, F^*)  \to  (T^*TM,F^*_t ),\quad
F^*_t(x,p) =k(t)\,{h(p,p)}+(1-k(t))\,{g(p,p)},\,
\label{evolutionofgeometricstructures}
\end{align}
such that
\begin{align}
\, \lim_{t\rightarrow 1} k(t)=1.
\label{limiteofk}
\end{align}
and where the norms $\sqrt{|h(p,p)|}$ and $\sqrt{|g(p,p)|}$ are defined by
\begin{align}
h(p,p)=\sqrt{|h_{ij}p^ip^j|},\quad g(p,p)=\,\sqrt{|g_{ij}p^ip^j|}.
\end{align}

The {\it time inversion operation} $T_t$ is defined in local coordinates on $T^*TM$ by
\begin{align*}
T_t:T^*TM \to T^*TM, \quad\quad
 (u,p)=(x,y,p_x,p_y)\mapsto (T_t(u),T^*_t(p))=(x,-y,-p_x,p_y).
\end{align*}
The speed components $\beta_x$ and the acceleration components $\beta_y$ of the vector field $\beta\in \Gamma \,TTM$ are determined by the time inversion operator $T_t$,
\begin{align*}
\beta_x:=\frac{1}{2}(\beta\,-T_t(\beta)),\quad \beta_y:=\frac{1}{2}(\beta\,+T_t(\beta)).
\end{align*}
 The classical {\it Hamiltonian function} of a DCRS  is defined by
\begin{align}
{H}(u,p):=F^* (u,p)-F^*(T_t(u),T^*_t (p))=2{\beta}^{i}(u)p_{i},\quad i=1,..., 8N.
\label{randershamiltonian}
\end{align}
The Hamiltonian \eqref{randershamiltonian} corresponds to a {\it time orientation average} of the Cartan-Randers function associated with a particular form of classical Hamiltonian (see for instance \cite{Syngespecial1972}, p. 22). It is the first stage towards the {\it total averaged Hamiltonian}. Also, note that a Cartan-Randers metric is non-reversible: $F^* (u,p)\neq F^* (T_t(u),-p)$ for a general $p$.

The Hamilton equations for ${H}(u,p)$ are
\begin{align}
\dot{u}^i=\frac{\partial H(u,p)}{\partial p_i}=\,\beta^i(u),\quad \dot{p}_i=\,-\frac{\partial H(u,p)}{\partial u^i}=\,-2\frac{\partial \beta^k(u)}{\partial u^i}p_k,\quad i,j,k=1,...,8N.
\label{Hamiltonianequation}
\end{align}
where the time derivatives are  respect to the non-compact time parameter $\tau$. Note that the $t$-time parameter and the $\tau$-time parameter are essentially different.
\\
{\bf Example}.
In order to illustrate the relation between Cartan spaces and Hamiltonian formalism, let us consider an example investigated by J. Synge that ties Cartan-Randers spaces with the Hamiltonian model for relativistic point charged particles \eqref{randershamiltonian} (see \cite{Syngespecial1972}, {\it paragraph} $12$). It can be shown that the Hamiltonian constraint
\begin{align}
\bar{H}(p(\xi),\xi)=\, \eta^{\mu\nu}_4(p_{\mu}+\bar{\beta}_{\mu})(p_{\nu}+\bar{\beta}_{\nu})+1 =0,
\end{align}
where the variables $(\xi,p)$ are conjugated and $\bar{H}(p(\xi),\xi)$ is the Hamiltonian function, has the same Hamilton equations than the Euler-Lagrange equations of the {\it Finsler-Randers functions}
\begin{align}
f_{\pm}(\xi,\dot{\xi})=\,\bar{\beta}_{\mu}\dot{\xi}^\mu \pm\, \eta^4_{\mu\nu}\dot{\xi}^\mu\,\dot{\xi}^\nu.
\end{align}

The Hamiltonian \eqref{randershamiltonian} corresponds to the {\it dual averaged Finsler function} $\langle f\rangle= \,\frac{1}{2}(f^*_{+}+\,f^*_{-})$ for $\bar{\beta}=\beta$ and is {\it the average  on the positive direction of time and negative direction of time of the energy function} $E(\xi,\dot{x})$ associated with $H(\xi,p)$.
This interpretation can be extended to DCRS in a way that  each molecule of the ideal classical gas is assumed to be composed by a pair of classical particles, one evolving in the positive  direction in the time parameter $t$ and the other evolving in the negative direction, such that the effective dynamical system is a pair of  molecules evolving in opposite directions in the parameter $t$.

For DCRS spaces, the non-degeneracy of the fundamental tensor $g$ of the underlying Cartan-Randers space is ensured if
 the vector field  ${\beta}$ is bounded by the metric $\eta^*$, $\|
\beta\|_{\eta^*} <1$. This implies in particular that all the {\it space-like components} of $\beta(x,y)$ are bounded,
\begin{align}
|\beta^i_{\vec{x}}|\leq c,\quad |\beta^i_{\vec{y}}|\leq a_m,\quad i=0,...,3N.
\label{boundaccelerationandspeed}
\end{align}
The time parameter  in these velocities is respect the non-compact time $t\in I$.
The same conditions imply that the function $F^*:\mathcal{C}\to R$ is positive. Moreover, that for the equivalence between Cartan spaces and Hamiltonian systems, the non-degeneracy of the metric is required.

If $( \mathcal{C},F^* )$ is a Cartan-Randers space
that evolves to the final {\it conic dual Riemannian structure}\footnote{The conic dual Riemannian structure $(TM,h )$ is only defined in a cone $\mathcal{C}$ of $T^*TM$.} $(TM,h )$ by a definite geometric evolution $U_t$ operation, for each value of $t$ 
there is a Cartan-Randers space $(TM,F^*_t)$. Applying the averaging operation to
\eqref{randershamiltonian} to $F^*_t$ one obtains the corresponding Hamiltonian of a DCRS,
\begin{align*}
H_t(u,p)& =\,F^*_t (u,p)-F^*_t(T_t(u),T^*_t (p))\\
& =\big((1-k(t))g^{ij}(u,p)p_ip_j+\,t\langle g^{ij}\rangle p_ip_j\big)\\
& -\big((1-k(t))g^{ij}(T_t(u),T^*_t(p))p_ip_j+\,t\langle g^{ij}\rangle p_ip_j\big)\\
& = (1-k(t))g^{ij}(u,p)p_ip_j\,-(1-k(t))g^{ij}(T_t(u),T^*_t(p))p_ip_j\\
& =\,(1-k(t))\beta^k(u)p_k.
\end{align*}
Then by the limit condition \eqref{limiteofk}, the {\it  equilibrium Hamiltonian} (or completely averaged Hamiltonian) of a DCRS is identically zero,
\begin{align}
\lim_{t\to 1}\,  H_t(u,p)=0.
\label{finalhamiltoniantevolution}
\end{align}
The condition \eqref{finalhamiltoniantevolution} can also be interpreted in a weaker form. First, one considers the quantized form of the Hamiltonian \eqref{randershamiltonian},
 \begin{align}
 \hat{H}_t(u,p):=\,\big(1-k(t)\big)\,{\beta}^{i}(\hat{u})\hat{p}_{i}+\,\hat{p}_{i}\,{\beta}^{i}(\hat{u}),\quad t\in[0,1]\,\quad i=1,..., 8N,
 \end{align}
 where canonical commutation relations $[\hat{u}^i,\hat{p}_j]=\,\delta^i_j$ are assumed.
The {\it quantum states} corresponding to the limit cycles correspond to are zero modes of the Hamiltonian at equilibrium,
\begin{align}
\big(\lim_{t\to 1}\,  \hat{H}_t(u,p)\big)\,|\psi\rangle=0,\quad |\psi\rangle\,\in \mathcal{H}_0.
\label{averagehamiltonianevolution}
\end{align}
$|\psi\rangle$ represents a {\it physical quantum state} and $\mathcal{H}_0$ is the vector space of zero modes of the Hamiltonian $\hat{H}_1$. The mechanism provides a finite effective Hamiltonian acting on physical quantum spaces and the action is invariant under diffeomorphism transformation, since the effective Hamiltonian acting on a physical state is zero. Moreover, the averaged description provides a nice solution to the un-boundeness problem in deterministic quantum mechanics.

Once the deterministic dynamical system reaches a limit cycle, remains in the limit cycle if it is not perturbed from the environment. This is a form of stating the Principle of Inertia for quantum states, since a system will continue in the same kinematical state if it is not act by the environment. If the limit cycle is a point, this corresponds to the classical Principle of Inertia.

The {\it natural tendency} towards and the persistence of the limit cycle could be argued by an statistical argument if the cyclic limit define the most probable state compatible with non-linear dynamicas towards a system composed by many particles evolves, as a generalization of a point limit cycle or equilibrium states. Moreover, the persistence on time of the limit cycle can be interpreted using the {\it ergodic hypothesis} \cite{ArnoldAvez}. If the system is ergodic, the phase averaging operation $\langle\cdot \rangle_{\mathcal{C}_u}$ is equivalent to the time averaging operation. Thus as a result, the state of the DCRS evolves on the $t$-time towards the zero average Hamiltonian by the condition \eqref{averagehamiltonianevolution} and the system  expends more time near zero mode of $H$ than in other states.

Following this interpretation, the limit cycle  state is {\it reached} faster for systems composed by a large number of particles. This is because the phase averaged state is reached faster for systems with a large number of degrees of freedom, since for such systems it will be more frequent to pass through each of the possible values. Moreover, in order to make easier the applicability of the ergodic hypothesis one can restrict the dynamics to happen on compact domains of $M$. Thus DCRS systems are characterized by domains of compactness in $TM$. In this picture, it is remarkable that gravity is not put as the element of dissipation of information, in contrast with 't Hooft proposal.

Let us remark that a main difference between DCRS and 't Hooft's deterministic framework is on the notion of time. In DCRS, time is a two-dimensional parameter $(\tau,t)$. The parameter $\tau$ is the {\it external time} that we use to describe how a quantum state $|\psi\rangle$ changes when interacting with the environment. This time corresponds with the usual notion of {\it macroscopic time} used in quantum mechanics or field theory. The internal $t$-time in DCRS is used to describe the internal evolution of the Hamiltonian $H_t(u,p)$ at the Planck scale and appears in the formalism as the parameter of an homotopy of Cartan-Randers structures. In contrast, the notion of time used in 'Hooft's framework is the usual notion of time as it is used in quantum mechanics, except for the fact that is a discrete parameter.

\section{The isometry group of a DCRS}
\
Given a DCRS specified by the metric $\eta\in\,\Gamma \,T^{(0,2)}TM$ and the vector field $\beta\in\,\Gamma \,TTM$ and a cone $\mathcal{C}\subset\,T^*TM$, an isometry of $F^*=\,\alpha+\,\beta$ is a diffeomorphism  $\phi:TM \to TM$ that preserves the Cartan-Randers function $F^*:\mathcal{C}\to M$,
\begin{align*}
F^*(\phi(u),\phi^*(p))=\,F^*(u,p).
\end{align*}
The isometry transformation must leave invariant each of the terms $\alpha$ and $\beta$. This is a consequence of the following algebraic relations,
\begin{align*}
\alpha(u,p)=\,\frac{1}{2}\big(F^*(u,p)+\,F^*(u,-p)\big),\quad \beta(u,p)=\,\frac{1}{2}\big(F^*(u,p)-\,F^*(u,-p)\big),\,\quad \forall\, p\in T^*_uTM,\,u\in\,TM.
\end{align*}
Thus for the isometry $\phi :TM\to TM$, the following relation hold,
\begin{align*}
\sqrt{-\eta^{ij}(u)p_ip_j}=\sqrt{-\eta^{ij}(\phi(u))(\phi^*{p})_i(\phi^*{p})_j},\quad\quad \beta^ip_i=\,\beta^i(\phi(u))(\phi^*p)_i.
\end{align*}
Therefore, the isometry $\phi$ must left invariant the metric $\eta$.
The metric $\eta$ is the direct sum $\eta=\sum^N_{k=1}\, \oplus \,\eta_S(k)$ and each isometry of $\eta$ acts independently for each value of $k$.

 The metric $\eta$ can be seen as a gauge field  that associates to each degree of freedom $\{k=1,...,N\}$ a copy of the Sasaki metric $\eta_S(k)$ on each $TM_4$. Thus the
isometries of $\eta$ are determined by the {\it isometries} $\phi(k)$ of each metric $\eta_S(k)$, that since it is a dual Sasaki type metric, are associated with the isometries of $F^*$ by dualizing the isometry group.
Therefore, the isometry group of the metric  $\eta$ is the direct sum
\begin{align}
Iso(\eta)=\,\sum^N_{k=1}\,\oplus  \, Iso(\eta_S(k)).
\end{align}
Since each of the Sasaki type metrics has the structure group isomorphic to 
\begin{align*}
Iso(\eta_S(k))\cong G\times G
\end{align*}
for some given Lie group $G$, the {\it isometry group} $Iso(\eta)$ must be of the form
\begin{align*}
Iso(\eta)\cong \,\sum^N_{k=1}\,\oplus \,Iso(\eta_S(k))=\,\sum^N_{k=1}\,\oplus\,G\times G,
\end{align*}
with $G$ a sub-group of $O(1,3)$. Therefore, the {\it gauge group} $G(k)\times G(k)$ is contained in the group $O(1,3)\times O(1,3)$, a relation that for each $k$ is
 \begin{align*}
 G(k)\times G(k)\subset \, \big(O(1,3)\times O(1,3)\big)|_k, \quad k=1,...,N.
 \end{align*}
As a consequence one has the relations
\begin{align}
Iso(\eta)\cong \,\sum^N_{k=1}\,\oplus\,Iso(\eta_S(k))\subset \,\sum^N_{k=1}\,\oplus\,\big(O(1,3)\times O(1,3)\big)|_k.
\label{structureofthegaugegroup}
\end{align}
Moreover, the {\it phases} $\theta(k)\in\, \big(O(1,3)\times O(1,3)\big)_k$ are defined independently  for each $k\in\{1,...,N\}$.

On the other hand, the isometries of $F^*$ must left invariant the vector $\beta(u)\in \,T_uTM$. In a similar way as before, we note that the vector field $\beta$ is decomposed as
\begin{align}
\beta=\,\sum^N_{k=1}\,\beta(k)=\,\sum^N_{k=1}\,\oplus\,\beta_x(k)\oplus \beta_y(k),
\end{align}
which shows the independence on the evolution equations for the $x$ and $y$ coordinates. The action of the isometry group is determined by the actions
\begin{align}
\theta_k:\big(O(1,3)\times O(1,3)\big)_k\,\times \,T_{(x,y)}TM\to T_{(x,y)}TM,\quad ((\theta_x,\theta_y),(\beta_x,\beta_y))\mapsto (\theta_x\cdot\beta_x,\theta_y\cdot\beta_y),
\end{align}
where the actions $\theta_x\cdot \beta_x$ and $\theta_y\cdot \beta_y$ are the standard vector representations of the group $G$.

For the sake of simplicity, let us assume that $\eta_4$ is the Minkowski flat metric. Then $G\cong O(1,3)$.
The invariance of the vector field $\beta\in \Gamma TTM$ under isometries is equivalent to the invariance  of each jet $(\beta_x(k),\beta_y(k))\in J^2_0(\xi)$  for the fundamental degree of freedom labeled by $k\in\{1,...,N\}$.
The four vector $\beta_x(k)\cong\dot{\xi}\in $ is assumed to be timelike or lightlike and the four acceleration $\beta_y(k)\cong \ddot{\xi}\in\,J^2_0(\xi)$ is spacelike for each $k$. As we assume that the metric $\eta_4$ is the Minkowski metric, the maximal sub-group of $O(1,3)\times O(1,3)$ leaving invariant the timelike vector velocity and the spacelike vector acceleration $(\beta_x(k),\beta_y(k))=(\dot{\xi},\ddot{\xi})$ is the abelian group of translations in the plane $T(2)\cong U(1)\times U(1)$. Thus, the symmetry of a DCRS can be described by a model with {\it fields} defined over the denumerable set $\{1,...,N\}$, and the isometries leaving invariant an state are {\it fibers} at each point $u\in M$ isomorphic to the Lie group $U(1)\times U(1)$.

The above construction can be extended to define local gauge fields defined on a compact subset $K$ of  $M_4$. Let us consider an arbitrary  point $\xi\in \,K\subset M_4$ and a macroscopic observer, described as a timelike vector field $W\in\,\Gamma TM_4$. Each manifold  $M^k_4$ is not only diffeomorphic to the model manifold $M_4$, but also isometric. Thus, one can use the isometries to construct the associated  world lines to the {molecules} $\{1,...,N\}$ as curves in the same $M_4$. There will be one of such curves $\gamma(\xi)$ that is the closest to $\xi$.  The distance is calculated using the Euclidean metric $\bar{\eta}_4(W)$ associated to the observer $W$. Let us assume that such distance is realized  for the molecule denoted by the integer $k_1$. Then at the point $\xi\in M_4$  we can define the corresponding {\it local gauge rotation} as the one associated with the symmetry transformation in $U(1)\times U(1)$ acting on $(\beta_x(k_1),\beta_y(k_1))$ for the closer position $\beta_x(k_1)(t_0)$ to $\xi$. Therefore, at the point $\xi$, there is defined an unique phase transformation (except for measure zero sets in $M_4$, which are equidistant from two different particles $k_1$ and $k_2$, or the points corresponding to collisions). Moreover, note that the assignment of a phase to the point $\xi$ does not depend on the vector field $W$, since although the distance defined by $\bar{\eta}_4(W)$ depends on $W$, the curve $\gamma(\xi)$ depends only on $\xi$.

In this way, there is a mechanism to define {\it sections} of a {\it sheaf}  $\pi:\mathcal{S}\to M_4$, where the fibers of the sheaf are diffeomorphic to the product group $U(1)\times U(1)$.
The sections of $\mathcal{S}$ introduced in such way are in general discontinuous.  A natural way to control the discontinuity of the sections is to have  control of the initial differences between the $U(1)\times U(1)$ phases along the world line of the molecules $\{1,...,N\}$. For this it is needed a control on the world-line of the molecules $\{1,...,N\}$ and this is provided by the bounds on the acceleration and speed of the world-lines and should be completely determined by the regularity properties of the functions $\{\beta^i:TM\to R,\,i=1,...,N\}$ in the Hamiltonian \eqref{randershamiltonian}.

\section{Effective actions for DCRS}

\
For classical  Yang-Mills models there is a natural {\it dualization} of the gauge symmetry \cite{ChanTsou, ChanFaridaniTsou}. In the case of an abelian $U(1)$ {\it pure gauge theory}, the Lagrangian has a dualized symmetry, $U(1)\times U(1)$. Moreover, the fundamental lagrangian in classical and quantum electrodynamics are Lorentz invariant. These symmetries are exactly the ones that DCRS have. This suggests that a natural way to relate the dynamics of DCRS described by a Hamiltonian \eqref{randershamiltonian} with phenomenological fields as they appear in field theory, living in the four dimensional model manifold $M_4$, is through  an effective relativistic, $U(1)\times U(1)$-gauge invariant model.

\subsection*{Information loss process in DCRS and effective variables}

Let us consider diffeomorphisms of the form
\begin{align*}
&J:{R}^{8N}\to R^{8N},\\
&(x^0(k),x^1(k),x^2(k),x^3({k}),\dot{x}^0(k),\dot{x}^1(k),\dot{x}^2(k),\dot{x}^3(k))\\
&\mapsto\, (x^0(k),x^1(k),x^2(k),x^3({k}),A^0(x(k),\dot{x}(k)),A^1(x(k),\dot{x}(k)),A^2(x(k),\dot{x}(k)),A^3(x(k),\dot{x}(k))),
\end{align*}
with $k=1,...,N$. $A(x(k),\dot{x}(k))$ determines a field living on a copy of the tangent bundle $TM_4$, for each value of $k$; once reached the equilibrium at $t=1$, the Hamiltonian is effectively $H_1=0$. Then there is necessarily a point $(\xi,\dot{\xi})\in\,TM_4$ such that the value of the smooth function $A^\mu(x(k),\dot{x}(k))$ takes an stable value that persists on the time parameter $t$. Furthermore, for systems  with a large number of degrees of freedom, the equilibrium value is reached very fast and it is {\it stable}, except for changes originated in the environment, that produce the unitary evolution in the ordinary macroscopic time.

 The gauge potentials $A(\xi,\varphi)$ with dependence on the velocity or higher derivatives are  of the kind of {\it higher order jet fields} discussed in \cite{Ricardo12} and since they appear up to first derivatives, $A(\xi,\varphi)$ corresponds to a  {\it Finslerian gauge potential}.  The effective macroscopic description of the $1$-form potential associated with $A$ is obtained by {\it fiber integration},
\begin{align}
\langle A_\mu \rangle (x)=\int_{T_{\xi}M_4}\, |\psi(\xi, \varphi)|^2 A_\mu (\xi,\varphi) \,dvol(\varphi),\quad \mu=0,...,3,
\label{kineticaverage}
\end{align}
with $A_{\mu}(\xi,\varphi)\,=\eta^{(4)}_{\mu\nu}A^{\nu},$ with $\eta^{(4)}_{\mu\nu}(\xi)$ the components of the metric $\eta_4$.
This definition of the macroscopic gauge potential assumes a kinetic theory interpretation of $(|\psi(\xi,\varphi)|^2$ as the one particle distribution function in phase space. It also assumes that the integral is finite, even if performed in a non-compact space.

 Therefore, one can associate effective dynamical models to the effective degrees of freedom $(\xi,\dot{\xi})$ and $A$. The models have by construction the same gauge symmetry $U(1)\times U(1)$. They are also Lorentz invariant, since by construction the effective degrees of freedom  coincide with the dynamics of a particular molecule $k_0$ of the DCRS (by the ergodicity properties of the DCRS) and such degrees of freedom have bounded velocity.
The exterior derivative of $\langle A\rangle$ defines the {\it macroscopic electromagnetic field},
 \begin{align}
 f_{\mu\nu}(\xi)=\partial_\mu  \langle A\rangle_\nu(\xi)-\,\partial_\nu  \langle A\rangle_\mu(\xi).
 \label{electromagnetic field}
   \end{align}
Let us denote by $\star_4 $ the star operator of the Minkowski metric $\eta_4$.
Then we assume that the action for $\langle A\rangle$ is the {\it free action}
\begin{align}
\mathcal{A}_0=\,-\frac{1}{4\pi}\,\int_{M_4}\,d^4\xi\,(\star_4 f)_{\mu\nu}(\star_4 f)^{\mu\nu}.
\label{freeactionforf}
\end{align}
This is the simplest action which is invariant by the dual group $U(1)\times U(1)$.

There are other morphisms $J:R^{8N}\to R^{8N}$ that retain the symmetries of the DCRS. They should correspond to $U(1)$-gauge models with matters. However, the fundamental requirement is that the $U(1)\times U(1)$-gauge symmetry and the Lorentz symmetry must be preserved.

\section{Discussion}

The
relation between Cartan structures and Hamiltonian systems is  known for long time. In particular, that the geodesic equations of a Cartan structure are the Hamilton equations of an associated Hamiltonian function and viceversa was noted previously by Synge (see for instance \cite{Syngespecial1972}). Indeed, Synge works the example related with a Randers type structure, showing its relation with the corresponding Hamiltonian theory. We have shown in \cite{Ricardo06} that Hamiltonian functions linear on the momentum and constrained to have maximal speed and acceleration  correspond  to DCRS and can be interpreted as  {\it the a total averaged model} of a deterministic Cartan-Randers systems. By total averaging one means to average the Hamiltonian function in the positive and negative directions of the $t$-evolution and after to perform  the phase average. By the ergodic hypothesis, the phase average corresponds to time average.

Therefore, the use of Cartan-Randers metrics in the contest of
DCRS is quite different from the application of Randers metrics
as geometrization of the dynamics of point charged particles in external
electromagnetic fields (see for instance \cite{Miron06}). Indeed, our motivation to use DCRS  has a strong analogy with the motivation of the original work of G. Randers as an attempt
to define {\it non-reversible} space-times \cite{Randers}. In
DCRS there is an intrinsic, non-reversible dynamics in a generalized cotangent bundle, which is the
responsible for the emergence of the quantum states as limit cycle states of systems with a very large number of molecules. The phenomenological quantum state itself {\it evolves along an external time} $\tau\in\,R$ which is non-compact and essentially a different parameter than the parameter $t$ guiding the irreversible evolution leading to limit cycles states of several kind. The phenomenological evolution is by definition reversible, in contrast with the $t$-evolution that is non-reversible. It is this notion of {\it double time} $(t,\tau)\in \, [0,1]\times\,R$  which is beneath the notion of quantum mechanics as geometric evolution. Thus, to speak of a measurement at an instant, one needs to specify both parameters $(t,\tau)$; if only $\tau$ is specified, an intrinsic non-local description for quantum systems emerges.

As we mention, the physical interpretation of the degrees of freedom of a DCRS as molecules of a classical gas and that the number $N$ is rather large compared with the dimensions of the model manifold presuppose that such degrees of freedom do not correspond to Standard Model particles, that on the other hand are seen as phenomenological low energy descriptions. In view of this problematic state, it is certainly a good strategy to observe which known Standard Model matter can be described as condensates of molecules emerging from DCRS. It is interesting to note that formally all dynamical systems admitting a linear Hamiltonian looks like DCRS, except for the boundeness conditions. Thus, we expect to use the findings of 't Hooft to reduce some simple examples of field theories to DCRS.

Two direct consequences of DCRS framework are highlighted: an statistical interpretation of the Principle of Inertia, that emerges from the beneath dynamical structure of DCRS framework and the emergence of diffeomorphism invariance for the phenomenological models, that is represented by an effective zero classical Hamiltonian or by zero modes of a quantum Hamiltonian operator.

A more specific consequence from DCRS is that the accelerations vectors $\ddot{\vec{\xi}}$ are bounded for each degree of freedom. The natural maximal acceleration associated with the Planck length scale is of order $10^{52} m/s^2$. Despite this enormous acceleration is unlikely to be measured directly in a laboratory, the existence of a maximal acceleration have relevant implications for the absence of space-time singularities (see for instance \cite{Caianiello, RovelliVidotto}), with possible consequences for cosmology.  We have also seen that under some regularity conditions on the vector field $\beta$, the classical Hamiltonian \eqref{randershamiltonian} is bounded when acting on physical states, solving one of the difficulties in 't Hooft's theory in a natural way. This solutions do not call for a fundamental role of gravity, that could be indeed an emergent phenomena. It appeals to an universal statistical argument, as it is the tendency to the limit cycle state for systems with a large number of degrees of freedom.

 A DCRS has a large automorphism group. Such group can be effectively described as a dualized abelian gauge symmetry with gauge group $U(1)\times U(1)$. The emergence of the dualized abelian gauge group is related with the specific properties of the DCRS and with the fact that the dynamics is dissipative in the sense of existence of limit cycle states towards the system evolves. Note that the gauge fields obtained are of Finsler type, that makes natural to think on Finsler gravity as a phenomenological description of gravity. However, Finsler gravity implies violations of the Einstein's equivalence principle generically, since the spacetime structure in a Finsler spacetime is in general not metrizable.

 Among the natural generalizations of DCRS as described in this paper, let us mention the possibility of analogous constructions with higher dimensional model space-time manifolds $M_d$ with $d\geq 4$. One expect to relate such models with {\it emergent Yang-Mills gauge theories}, where the gauge symmetry is related with the isometry group of the manifold extra-dimensions via Kaluza-Klein mechanism. Again, the dualization of the isommetry group is expected, as well as the Lorentz covariance of the effective model if the interpretation of the spacetime is a four dimensional manifold $M_4$. Thus, the natural structure for $M_d$ is a product $M_d\cong M_4\times \,\tilde{K}$, with $K$ a compact Lie group. Therefore, the isometry group will be of the form $Iso(K)\times U(1)$ dualized, since it is the group leaving invariant both, the four dimensional metric $\eta_4$ and the vector field $\beta\in \,\Gamma T^*TM$.

One can also consider spinor representations of the Lorentz group associated with the metric $\eta_4$. Indeed, some of such spinor degrees of freedom can be described as deterministic systems (see for instance \cite{Hooft}).

  Finally, in view of the discussion presented in this work it is natural to wonder whether gravity is  an emergent phenomena too. There are some clues pointing on this direction, in particular if gravity is a macroscopic effect as discussed by several authors (see for instance \cite{Jacobson} and \cite{Verlinde}). Indeed, one expects that the weak equivalence principle has an emergent interpretation from systems with large number of degrees of freedom. If this is true, the weak equivalence principle will be gradually violated for systems with dynamics is on higher energy scale and eventually will be completely violated at the Planck scale for systems composed by individual molecules in DCRS.

\bigskip

{\bf Acknowledgement.} This version of the work was partial financially supported by The Riemann Center for Geometry and Physics, Leibnitz University Hannover and by a PNPD fellowship by CAPES (Brazil).

\end{document}